\documentclass[journal,9pt]{article}

\usepackage{arxiv}

\usepackage[utf8]{inputenc} 
\usepackage[T1]{fontenc}    
\usepackage{url}            
\usepackage{booktabs}       
\usepackage{amsfonts}       
\usepackage{nicefrac}       
\usepackage{microtype}      

\usepackage{pgfplots}
\usetikzlibrary{positioning,shapes,arrows,patterns,fit,backgrounds,calc,svg.path,3d,decorations.text,shapes.arrows,spy}
\usepackage{tikz,bm,color,pgfkeys,siunitx,ifthen}
\usepackage{todonotes}
\usepackage{graphicx}
\usepackage{subfigure}
\usepackage{amsmath,amssymb}
\usepackage{mathtools,nccmath}
\usepackage{array}
\usepackage{makecell}
\usepackage{caption}
\usepackage{multirow}
\usepackage{placeins}
\usepackage{amsmath}
\usepackage{enumitem}
\usepackage{algorithm}
\usepackage{algpseudocode}
 
\DeclareMathSymbol{\shortminus}{\mathbin}{AMSa}{"39}
\DeclareMathOperator*{\argmax}{arg\,max}

\DeclareMathOperator*{\softmax}{softmax}

\pgfdeclarelayer{background}
\pgfdeclarelayer{front}
\pgfsetlayers{background,main,front}
\makeatletter
\pgfkeys{%
	/tikz/on layer/.code={
		\pgfonlayer{#1}\begingroup
		\aftergroup\endpgfonlayer
		\aftergroup\endgroup
	},
	/tikz/node on layer/.code={
		\gdef\node@@on@layer{%
			\setbox\tikz@tempbox=\hbox\bgroup\pgfonlayer{#1}\unhbox\tikz@tempbox\endpgfonlayer\egroup}
		\aftergroup\node@on@layer
	},
	/tikz/end node on layer/.code={
		\endpgfonlayer\endgroup\endgroup
	}
}

\def\node@on@layer{\aftergroup\node@@on@layer}

\pgfplotsset{ every non boxed x axis/.append style={x axis line style=-},
	every non boxed y axis/.append style={y axis line style=-}}

\def\BibTeX{{\rm B\kern-.05em{\sc i\kern-.025em b}\kern-.08em
		T\kern-.1667em\lower.7ex\hbox{E}\kern-.125emX}}

\title{Self-attention for Enhanced OAMP Detection in MIMO Systems}
\author{Alexander Fuchs \\
Signal Processing and Speech Communication Lab.\\ Graz University of Technology\\
\And  Christian Knoll \\
Signal Processing and Speech Communication Lab.\\ Graz University of Technology\\
\And Nima N. Moghadam\\
Huawei Technologies Sweden AB
\And Alexey Pak\\
Huawei Technologies Sweden AB
\And Jinliang Huang\\
Huawei Technologies Sweden AB
\And Erik Leitinger\\
Signal Processing and Speech Communication Lab.\\ Graz University of Technology\\
\And Franz Pernkopf\\
Signal Processing and Speech Communication Lab.\\ Graz University of Technology\\
}

\begin{document}

\maketitle
\begin{abstract}
    
    Multiple-Input Multiple-Output (MIMO) systems are essential for wireless communications. Since classical algorithms for symbol detection in MIMO setups require large computational resources or provide poor results, data-driven algorithms are becoming more popular. Most of the proposed algorithms, however, introduce approximations leading to degraded performance for realistic MIMO systems. In this paper, we introduce a neural-enhanced hybrid model, augmenting the analytic backbone algorithm with state-of-the-art neural network components. In particular, we introduce a self-attention model for the enhancement of the iterative Orthogonal Approximate Message Passing (OAMP)-based decoding algorithm. In our experiments, we show that the proposed model can outperform existing data-driven approaches for OAMP while having improved generalization to other SNR values at limited computational overhead.
    
\end{abstract}
\section{Introduction}\label{sec:introduction}
Multiple-Input Multiple-Output (MIMO) systems have become an integral part of communication standards over the past few years. They offer increased spectral and energy efficiency, as well as improved spatial resolution between users, and link reliability~\cite{goldsmith2003capacity,larsson2014massive}. Efficient and precise channel estimation and symbol detection is vital to leverage the benefits of MIMO systems. 
Although the Maximum Likelihood (ML) detector can achieve optimal performance for the detection problem, it scales exponentially with the size of the MIMO system, which limits its practical relevance.
The sphere-decoding algorithm alleviates this by limiting the search space but remains intractable for large systems~\cite{guo2006algorithm}.
The Zero Forcing (ZF) detector and Linear Minimum Mean Squared Error (MMSE) detector~\cite{RusekSPM2013}, are able to solve the detection problem efficiently but do not perform as well as the ML detector.

There are two alternative directions for improving the detection quality at limited computational complexity.
First, approximating the maximum likelihood solution by iterative message passing algorithms: these include Expectation Propagation (EP), Approximate Message Passing (AMP), and Orthogonal AMP (OAMP)~\cite{cespedes2014expectation,donoho2009message,ma2017orthogonal}. Particularly the OAMP detector shows excellent performance for idealized independent and identically distributed (i.i.d.) input data and uncorrelated channels. Because of the implicit assumptions and approximations, the performance of the OAMP algorithm degrades for more realistic scenarios with correlated data and channels though.
Second, spurred by the recent successes of machine learning, it is tempting to attempt learning the detection directly from data. However, given the complexity of the problem, this would require huge amounts of training data and cannot be expected to generalize well to scenarios not captured in the training data.
Instead, successful data-driven approaches in such a complex setting, need to exploit neural networks and be build upon existing iterative frameworks via algorithm unrolling~\cite{9363511}. The first approach of this kind was DetNet~\cite{samuel2017deep}; later approaches that incorporate some data dependence -- such as MMNet and OAMPNet -- could improve the detection performance, especially for uncorrelated channels~\cite{khani2020adaptive,he2018model,he2020model}.
All these data-driven approaches learn relatively few static parameters, and although this accelerates the convergence and improves the detection performance, it still leaves room for improvement by fully utilizing the flexibility and capabilities of neural networks.
So-called neural-enhanced or hybrid models have recently been proposed with great success for Belief Propagation (BP) based detection~\cite{satorras2021neural}. These hybrid models rely on graph neural networks to refine the message updates within the BP iteration. The overarching goal is to combine the flexibility of neural networks with the modeling capabilities of generative models to:
(i) mitigate the approximation error of the generative model, (ii) accelerate convergence, and (iii) improve the detection performance.

If we want to achieve good detection performance reliably with good generalization (i.e., also for scenarios not seen during training), we need a close integration of purely data-driven approaches with generative models that capture strong prior information about the underlying system in the form of inductive biases.

In this paper, we propose a \emph{neural-enhanced} OAMP algorithm.
More precisely, we use a self-attention model for augmenting the OAMP iteration and, in doing so, reducing the dependence on the rigid assumptions of i.i.d. data and uncorrelated channels. 
Self-attention was developed for transformer models~\cite{vaswani2017attention}, which are now used in a wide range of areas ranging from speech, over image, to signal-processing applications~\cite{baevski2020wav2vec,dosovitskiy2020image,pinto2022can}.
One of the main benefits of using a self-attention model is that it is agnostic to the number of users and permutations of the transmitted signal. This allows for a symbiotic relationship with the analytical OAMP backbone algorithm that not only improves the decoding performance but also offers excellent generalization capabilities.\footnote{Code: https://github.com/alexf1991/self\_attention\_oamp\_mimo}
In our experiments, we analyze the model performance and generalization capabilities on a Rayleigh channel and evaluate the symbol detection performance on more realistic 3GPP channel data. 


\section{Self-attention enhanced OAMP for MIMO decoding}\label{sec:mimo_decoding}
The MIMO decoding setup for $M$ receive and $N$ transmit antennas is given by a linear system, i.e.,
\begin{equation}\label{eq:receivedSignal}
\small
    \mathbf{y}=\mathbf{H}\mathbf{x}+\boldsymbol{\nu},
\end{equation}
where $\mathbf{y}\in \mathbb{C}^M$ represents the received signal vector, $\mathbf{H} \in \mathbb{C}^{M\times N}$ is the channel matrix, $\mathbf{x} \in \mathbb{C}^N$ is a vector containing the transmit symbols, and $\boldsymbol{\nu} \in \mathbb{C}^M$ is a noise vector containing samples from a complex Gaussian distribution $\mathcal{CN}(0,\sigma)$ with standard deviation $\sigma$. 
In this work, we assume that $\mathbf{H}$ and $\sigma$ are known or estimated by a preceding channel estimation step.
The decoder has to determine an inverse solution of Eq.~\ref{eq:receivedSignal} by detecting the correct transmitted symbols $\mathbf{x}$.
The optimal estimate $\hat{\mathbf{x}}$ of the transmit symbols $\mathbf{x}$ is given by the maximum likelihood estimator, i.e.,
\begin{equation}
\small
    \hat{\mathbf{x}} = \argmax_{\mathbf{x}}  \mathcal{P}(\mathbf{x}|\mathbf{y},\mathbf{H})\label{eq:estimator}
\end{equation}
with
\begin{equation}
\small
    \mathcal{P}(\mathbf{x}|\mathbf{y},\mathbf{H}) = \dfrac{\mathcal{P}(\mathbf{y}|\mathbf{x},\mathbf{H})\mathcal{P}(\mathbf{x})}{\mathcal{P}(\mathbf{y}|\mathbf{H})}.
\end{equation}
Here, the data evidence $\mathcal{P}(\mathbf{y}|\mathbf{H})$ is obtained by marginalization over $\mathbf{x}$,
\begin{equation}
\small
    \mathcal{P}(\mathbf{y}|\mathbf{H}) = \sum_{\mathbf{x}\in \mathcal{S}} \mathcal{P}(\mathbf{y}|\mathbf{x},\mathbf{H})\mathcal{P}(\mathbf{x}).\label{eq:data_evidence}
\end{equation}
However, the estimate $\hat{\mathbf{x}}$ is in general computationally intractable as it involves the evaluation of  high-dimensional integrals (see Eq.~\ref{eq:estimator} and Eq.~\ref{eq:data_evidence})~\cite{RusekSPM2013}.
Therefore, several algorithms have been proposed to improve computational efficiency by approximating $\mathcal{P}(\mathbf{x}|\mathbf{y},\mathbf{H})$~\cite{donoho2009message,ma2017orthogonal,kirkelund2010variational}. OAMP-based algorithms have shown to perform well for symbol detection. They decouple the posterior pdf by assuming independent Gaussian distributed symbols $\mathbf{x}$. This leads to an iterative state evolution algorithm that, if converged, has the potential to retrieve the optimal symbol estimates $\hat{\mathbf{x}}$ for independent and identically distributed data and uncorrelated channels. The OAMP algorithm is defined in Algorithm \ref{alg:oamp}.
\begin{algorithm}
\small
\caption{OAMP algorithm for MIMO detection}\label{alg:oamp}
\hspace*{\algorithmicindent} \textbf{Input:} Channel matrix $\mathbf{H}$, received signal $\mathbf{y}$ and noise \newline \hspace*{\algorithmicindent} standard deviation $\sigma$, number of iterations $T$, \newline \hspace*{\algorithmicindent} number of receive antennas $M$, number of symbols $N$.\newline
\hspace*{\algorithmicindent} \textbf{Output:} Estimated symbols $\mathbf{x}_T$.

\hspace*{\algorithmicindent} \textbf{Initialize:} $\tau_{t-1} \gets 1$, $\mathbf{x}_{t-1} \gets 0$, $v_{t-1} \gets 1$
\begin{algorithmic}
    
\For{$t < T$}
    \State\vspace*{-\baselineskip}
    \begin{fleqn}[\dimexpr\leftmargini-\labelsep]
        \setlength\belowdisplayskip{0pt}
        \begin{equation}
                \hat{\mathbf{W}}_t \gets v_{t-1}^2\mathbf{H}^T\left(v_{t-1}^2 \mathbf{H}\mathbf{H}^T +\frac{\sigma^2}{2} \mathbf{I}\right)^{-1}\label{eq:W_hat}
        \end{equation}
    \end{fleqn}
    
    \State\vspace*{-\baselineskip}
    \begin{fleqn}[\dimexpr\leftmargini-\labelsep]
        \setlength\belowdisplayskip{0pt}
        \begin{equation}
                \mathbf{W}_t \gets \dfrac{2N}{\mathrm{tr}(\hat{\mathbf{W}}_t\mathbf{H})}\label{eq:W_norm}
        \end{equation}
    \end{fleqn}
        
    \State\vspace*{-\baselineskip}
    \begin{fleqn}[\dimexpr\leftmargini-\labelsep]
        \setlength\belowdisplayskip{0pt}
        \begin{equation}
            \mathbf{r}_t \gets \mathbf{x}_{t-1} + \mathbf{W}_t(\mathbf{y}-\mathbf{H}\mathbf{x}_{t-1})\label{eq:r_update}
        \end{equation}
    \end{fleqn}
    
    \State\vspace*{-\baselineskip}
    \begin{fleqn}[\dimexpr\leftmargini-\labelsep]
        \setlength\belowdisplayskip{0pt}
        \begin{equation}
        \mathbf{x}_{t} \gets \mathbb{E}\{\mathbf{s}|\mathbf{r}_t,\tau_{t-1}\} \label{eq:estimator_algo}
        \end{equation}
    \end{fleqn}
    
    \State\vspace*{-\baselineskip}
    \begin{fleqn}[\dimexpr\leftmargini-\labelsep]
        \setlength\belowdisplayskip{0pt}
        \begin{equation}
        v^2_t\gets \dfrac{||\mathbf{y}-\mathbf{H}\mathbf{x}_t||^2_2-M\sigma^2}{\mathrm{tr}(\mathbf{H}^T\mathbf{H})}
        \end{equation}\label{eq:v_calc}
    \end{fleqn}
    
    \State\vspace*{-\baselineskip}
    \begin{fleqn}[\dimexpr\leftmargini-\labelsep]
        \setlength\belowdisplayskip{0pt}
       \begin{equation}
        \mathbf{B} \gets \mathbf{I}-\mathbf{W}_t\mathbf{H} \label{eq:b_calc}
        \end{equation}
    \end{fleqn}
    
    \State\vspace*{-\baselineskip}
    \begin{fleqn}[\dimexpr\leftmargini-\labelsep]
        \setlength\belowdisplayskip{0pt}
       \begin{equation}
        \tau^2_t \gets \frac{1}{2N}\mathrm{tr}(\mathbf{B}\mathbf{B}^T)v_t^2 + \frac{1}{4N}\mathrm{tr}(\mathbf{W}_t\mathbf{W}_t^T)\sigma^2\label{eq:tau_calc}
        \end{equation}
    \end{fleqn}
    \vspace{0.1 cm}
\EndFor

\end{algorithmic}\label{alg:oamp}
\end{algorithm}

The OAMPNet models are both based on the OAMP algorithm~\cite{he2018model,he2020model}, but include additional scaling parameters, which improve the convergence of the algorithm, leading to better results at the same number of iterations.
The OAMPNet2 model uses the trainable parameter $\gamma$ to scale the updates of the linear estimate $\mathbf{r}_t$, exchanging Eq.\ref{eq:r_update} for:
\begin{equation}
    \mathbf{r}_t \gets \mathbf{x}_{t-1} + \gamma_t \mathbf{W}_t(\mathbf{y}-\mathbf{H}\mathbf{x}_{t-1})
\end{equation}
and a second parameter $\theta$ as a scaling parameter for $\mathbf{W}$, resulting in a modified $\mathbf{B}$ in Eq. \ref{eq:b_calc}, and variance $\tau^2$ in Eq. \ref{eq:tau_calc} for the non-linear estimator:
\begin{equation}
    \begin{split}
                \mathbf{B} &\gets \mathbf{I}-\theta_t\mathbf{W}_t\mathbf{H},\\
                \tau^2_t &\gets \frac{1}{2N}\mathrm{tr}(\mathbf{B}\mathbf{B}^T)v_t^2 + \frac{\theta^2}{4N}\mathrm{tr}(\mathbf{W}_t\mathbf{W}_t^T)\sigma^2.
    \end{split}
\end{equation}
Furthermore, it includes the linear estimate $\mathbf{r}_t$ in addition to the non-linear estimate $\mathbb{E}\{\mathbf{s}|\mathbf{r}_t,\tau_t\}$ to create $\mathbf{x}_{t+1}$, using two additional trainable parameters $\phi_t$ and $\zeta_t$ as weighting factors, and the symbol alphabet $\mathcal{S}$ given by the modulation scheme. The updated symbol estimates are therefore given by
\begin{equation}
    \mathbf{x}_{t} = \phi_t(\mathbb{E}\{\mathbf{s}|\mathbf{r}_t,\tau_{t-1}\} - \zeta_t \mathbf{r}_t). \label{eq:oampnet2_est}
\end{equation}
OAMPNet2 performs well for i.i.d data and weakly correlated channels, however, the detection performance degrades significantly in more realistic scenarios characterized by stronger correlations between the individual MIMO channels as the approximations introduce a model mismatch. 
The goal of our proposed neural-enhanced hybrid model is to (i) reduce the impact of correlated channels on the detection performance by mitigating the model mismatch and (ii) to improve the convergence behavior of the algorithm. Both measures lead to improved detection performance.
Our proposed model is based on the OAMP algorithm and extends the algorithm with a data-driven self-attention model to augment the predictions.   

\subsection{Self-attention}\label{sec:multi_headed_att}
The concept of self-attention was first introduced for transformer models in natural language processing~\cite{vaswani2017attention}. Transformers are a special type of sequence-to-sequence DNNs, which were introduced to improve the contextual understanding of words. They implement a multi-stage encoder-decoder structure, where each stage includes a multi-headed attention layer, that uses self-attention to establish relation between words. Here, each word is represented by a vector token $\mathbf{z}_i$, where $i\in\{1,2\dots T\}$ and $T$ is the number of tokens.

The self-attention layer creates multiple projections of the tokens that are then used to create an attention matrix $\mathbf{A}$ which intends to represent the relations between tokens. This attention matrix is based on scalar products between the tokens and can adjust according to the input sentence.
To create these projections, three different linear layers are applied to every input token $\mathbf{z}_i$. Since each of the three resulting vectors fulfills a different role within the self-attention layer, they are referred to as:
\begin{itemize}[noitemsep]
    \item \emph{key} vector: $\mathbf{k}_i = \mathbf{K} \mathbf{z}_i$,
    \item \emph{query} vector: $\mathbf{q}_i = \mathbf{Q}\mathbf{z}_i$,
    \item \emph{value} vector: $\mathbf{v}_i = \mathbf{V}\mathbf{z}_i$,
\end{itemize}
where $\mathbf{K}$, $\mathbf{Q}$, and $\mathbf{V}$ are linear projection matrices, that are learned during training. 
This naming convention originates from the sequence-to-sequence literature, as there the decoder queries a key for a specific token from the encoder, or vice versa, to perform matching between tokens.
The intention of the \emph{key} and \emph{query} vectors are to form connections between tokens that can benefit from information exchange and to decouple tokens that are independent. The elements of the attention matrix $A_{i,j}$ are then created by first taking the absolute value of the scalar product between the key and query vectors of each input $A_{i,j}=|\mathbf{k}^{\mathrm{T}}_i \mathbf{q}_j|$, before a $\softmax$ function, using an inverse temperature parameter $\beta$, is applied:
\begin{equation}
\small
    \bar{A}_{i,j} = \softmax_j\left(\beta A_{ij}\right).
\end{equation}
Here, the inverse temperature is typically chosen as one over the square root of the number of used vector dimensions $\beta= 1/ \sqrt{d_k}$, where $d_k=\mathrm{dim}(\mathbf{k})$.
The normalized attention matrix $\bar{\mathbf{A}}$ is then applied to the value vector tokens $\mathbf{v}_j$ to create the output tokens via a weighted sum,
\begin{equation}
\small
    \mathbf{\bar{v}}_i = \sum_j \bar{A}_{i,j} \mathbf{v}_j. \label{eq:attention}
\end{equation}
Afterwards two layers $\mathbf{W}_1 \in \mathbb{R}^{d_k\times d_k},\mathbf{W}_2\mathbb{R}^{d_k\times d_k}$, layer normalization and a $\mathrm{ReLU}$ activation function are applied to create the final output vector tokens $\mathbf{\bar{z}}_i$ of the self-attention~\cite{vaswani2017attention}. 
Layer normalization (LN) for the latent vector $\mathbf{z}$ is defined as
\begin{equation}
\small
    \mathrm{LN}(\mathbf{z}) = \frac{\mathbf{z}-\hat{\mu}}{\hat{\sigma}}, \hat{\mu}=\frac{1}{C_l}\sum_c^{C_l} z_i,\ \hat{\sigma} = \sqrt{\frac{1}{C_l}\sum_c^{C_l}(z_i-\hat{\mu})^2},
\end{equation}
where $C_l$ is the total number of channels, i.e. vector dimensions $c$.  
Each layer produces an output of the same dimension as the previous layer to enable the use of shortcut connections and prevent vanishing gradients:
\begin{equation}
\begin{split}
        \mathbf{a}_i &= \mathrm{LN}(\mathbf{z}_i+\mathbf{\bar{v}}_i),\\
        \mathbf{b}_i &= \mathbf{W}_2\mathrm{ReLU}(\mathbf{W}_1\mathbf{a}_i),\\
        \mathbf{\bar{z}}_i &= \mathrm{LN}(\mathbf{b}_i+\mathbf{a}_i).
        \end{split}
\end{equation}
\subsection{Self-attention for MIMO detection}\label{sec:complex_valued_att}
For MIMO detection algorithms independence from the user order, and the total number of users is essential. Self-attention models can explicitly include these properties in the model if no position encodings are used. This helps the model to learn more efficiently and improve generalization, similar to the translation invariant kernel functions in Convolutional Neural Networks (CNNs).
In our proposed setup, the self-attention model acts as a post-processing step for the linear estimator of the OAMP iteration. Meaning, that the OAMP algorithm first generates estimates for $\mathbf{r}_t$ and $v^2_t$, which are then included in the input for the self-attention model. Here, we omit Eq.~\ref{eq:estimator_algo}, Eq.~\ref{eq:b_calc}, and Eq.~\ref{eq:tau_calc}. Instead, the model predicts symbol likelihood estimates $\mathcal{P}(s_k|f(\mathbf{z}_{i,t}))_t$ for each of the $K$ symbols $s_k$ in the alphabet defined via the modulation scheme, which are then used to calculate the expected symbol estimates.:
\begin{equation}
    \mathbf{x}_{t} \gets \mathbb{E}\{\mathbf{x}\in \mathcal{S}|\mathcal{P}(s_k|f(\mathbf{z}_{i,t}))_t\}.
\end{equation}
Additionally, the model also predicts a refined the variance estimate $\hat{v}^2_t=g(f(\mathbf{z}_{i,t}))$.
For appropriate data representation, we create two input vector tokens per user, representing real- and imaginary parts. We compose symbol and channel information for each user in a vector, and combine it with the current OAMP estimates for $\mathbf{r}_t$, $\boldsymbol{v}^2_t$, and the symbol likelihood estimates of the previous iteration $\mathcal{P}(s_k|f(\mathbf{z}_{i,t}))_{t-1}$. 
\subsubsection{Data preparation}\label{sec:input_prep}
In order to optimize the model for a MIMO detection task we structure the input data into $2N$ vector tokens, with two tokens for the real- and imaginary part of each user. This data structure preserves all information while staying limited in size and computational complexity, even for large MIMO systems. 
Each input vector token $\mathbf{z}_i$ consists of the following components:
\begin{itemize}[noitemsep]
    \item The current latent state of the transformer network in the form of a latent vector $\mathbf{\bar{b}}_{i,t-1} \in \mathbb{R}^{d_b}$, where $d_{\bar{b}}=dim(\mathbf{\bar{b}}_{i,t-1})$ is the number of used state channels.
    \item The received signal $\mathbf{y} \in \mathbb{R}^{M}$.
    \item The row of channel matrix corresponding to user $n$ $\mathbf{H}_n \in \mathbb{R}^{M}$.
    \item The last estimate for $\mathcal{P}(s_k|f(\mathbf{z}_{i,t}))_{t-1} \in \mathbb{R}^{K}$.
    \item The last received symbol estimate $x_{t-1} \in \mathbb{R}^{1} $.
    \item The current estimate for $r_t \in \mathbb{R}^{1}$.
    \item The current estimate for $\hat{v}^2_t \in \mathbb{R}^{1}$.
\end{itemize}
The final input vectors $\mathbf{z}_{i,t}$ are created by concatenating the vector components above along the channel dimension and normalizing using layer normalization. This results in a vector $\mathbf{z}_i \in R^{C}$, with $C=2M+K+d_{\bar{b}}+3$.
\subsubsection{Estimation}\label{sec:estimation}
The self-attention model $f(\mathbf{z}_{i,t})$ then processes the tokens into latent vectors $\mathbf{b}_{i,t}=f(\mathbf{z}_{i,t}) \in \mathbb{R}^{C}$, which are then used for the estimation of $\mathcal{P}(s_k|f(\mathbf{z}_{i,t}))_t$ and $\hat{v}^2_t$. The first $d_{\hat{b}}$ dimensions of $\mathbf{b}_{i,t}$ are also used as the new latent state $\mathbf{\bar{b}}_{i,t}$. For the likelihood estimation of the alphabet, two dense layers are applied to each $\mathbf{b}_{i,t}$:
\begin{equation}
\small
    \mathcal{P}(s_k|f(\mathbf{z}_{i,t}))_t = \sigma(\mathbf{W}_2 \mathrm{LN}(\mathrm{ReLU}(\mathbf{W}_1 \mathbf{b}_{i,t}))),
\end{equation}
where $\mathbf{W}_1\in \mathbb{R}^{C\times d_{\bar{b}}}$, $\mathbf{W}_2\in \mathbb{R}^{d_{\bar{b}}\times 1}$ are dense layers, $\mathrm{LN}$ is layer normalization, and $\sigma(\cdot)$ is the logistic sigmoid activation function. 
For the estimation of $\hat{v}^2_t$, we first average $\mathbf{b}_i$ over the tokens, and then apply a single dense layer $\mathbf{W}_3 \in \mathbb{R}^{d_{\bar{b}}\times 1}$ reducing the vector to a scalar. Then a squared logistic sigmoid activation function is applied to create $\hat{v}^2_t$: 
\begin{equation}
\small
    \hat{v}^2_t = \sigma\left(\mathbf{W}_3\left(\frac{1}{2N} \sum_i^{2N} \mathbf{b}_{i,t}\right)\right)^2.
\end{equation}
\section{Experiments}\label{sec:experiments}
In our experiments, we evaluate the performance and generalization behavior as well as the runtime of our proposed OAMP-Self-Attention (OAMP-SA) algorithm and compare it against OAMP, OAMPNet2, and the MMSE estimator without interference cancellation according to~\cite{RusekSPM2013}. Section~\ref{sec:generalization} provides a comparison of the generalization behavior of the different algorithms; in Section~\ref{sec:realistic_channel} we analyze the performance in a more realistic MIMO scenario using 3GPP channel matrices; and Section~\ref{sec:runtime_evaluation} analyzes the runtime for large MIMO systems. The data-driven models OAMP-SA and OAMPNet2 are trained using a cross-entropy loss for each iteration:
\begin{equation}
    \mathcal{L}_{CE} = \sum_t \sum_k -{(y_k\log(\mathcal{P}(s_k|f(\mathbf{z}_{i,t}))_t)+ (1 - y_k)\log(1 - \mathcal{P}(s_k|f(\mathbf{z}_{i,t}))_t))},
\end{equation}
where $y_k$ is a binary label for the specific symbol probability.
We perform the training of the model for 25 epochs using an ADAM optimizer with a learning rate of $lr=0.001$~\cite{kingma2014adam}. After 10 and 20 epochs we decay the learning rate by a factor of 10. All iterative algorithms are evaluated after 10 iterations.
\subsection{Model Generalization}\label{sec:generalization}
In this experiment, we investigated a MIMO system using $N_t=8$ transmit antennas and  $N_r=16$ receive antennas for a Gaussian i.i.d channel with Rayleigh fading using a QAM-4 encoding.
The datasets consist of 80k channel realizations for training, as well as a validation set, and test set with 20k channel realization, for each integer value within the SNR range of 5-14~dB.
To analyze generalization for the data-driven models, we evaluate two settings: (i) Training the models on MIMO data with a specific SNR and specific correlation coefficients $c_r$, $c_t$, and then evaluating on data with the same SNR and correlation, (i.e., the non-generalizing setting) and (ii) training the model on data with a fixed SNR value of 10~dB an with fixed correlation coefficients $c_r=c_t=0.1$, and then evaluating the same model over the entire SNR range and at multiple correlations (i.e., the generalizing setting). 
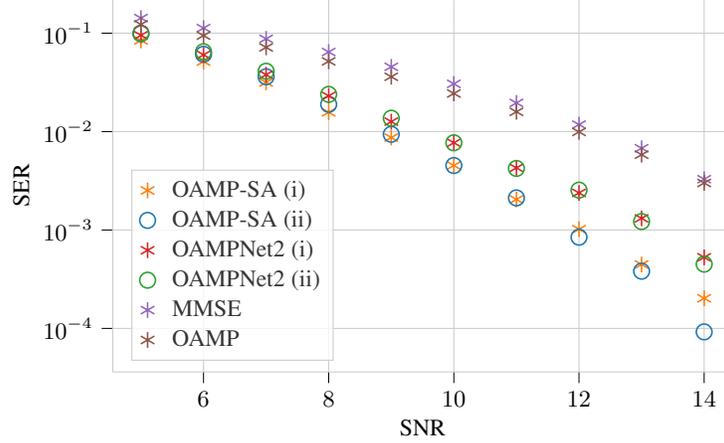
\begin{figure}[h]
	\centering
	\footnotesize
	\resizebox{0.6\linewidth}{!}{
\begin{tikzpicture}

\definecolor{color0}{rgb}{0.12156862745098,0.466666666666667,0.705882352941177}
\definecolor{color1}{rgb}{1,0.498039215686275,0.0549019607843137}
\definecolor{color2}{rgb}{0.172549019607843,0.627450980392157,0.172549019607843}
\definecolor{color3}{rgb}{0.83921568627451,0.152941176470588,0.156862745098039}
\definecolor{color4}{rgb}{0.580392156862745,0.403921568627451,0.741176470588235}
\definecolor{sienna1408675}{RGB}{140,86,75}
\begin{axis}[
legend cell align={left},
height=0.4\linewidth,
width=0.6\linewidth,
legend style={fill opacity=0.8, draw opacity=1, text opacity=1, at={(0.03,0.03)}, anchor=south west, draw=white!80!black},
log basis y={10},
axis line style={white!80!black},
tick align=outside,
tick pos=left,
x grid style={white!80!black},
xlabel={SNR},
xmajorgrids,
xmin=4.55, xmax=14.45,
xtick style={color=black},
y grid style={white!80!black},
ylabel={SER},
ymajorgrids,
ymin=3.61184339429159e-05, ymax=0.225404276644085,
ymode=log,
ytick style={color=black},
ytick={1e-06,1e-05,0.0001,0.001,0.01,0.1,1,10},
yticklabels={\(\displaystyle {10^{-6}}\),\(\displaystyle {10^{-5}}\),\(\displaystyle {10^{-4}}\),\(\displaystyle {10^{-3}}\),\(\displaystyle {10^{-2}}\),\(\displaystyle {10^{-1}}\),\(\displaystyle {10^{0}}\),\(\displaystyle {10^{1}}\)}
]
\addplot [semithick,color1, mark=asterisk, mark size=3, mark options={solid}, only marks]
table {%
5 0.0843268036842346
6 0.0514261275529861
7 0.0315871760249138
8 0.0157119166105986
9 0.00869939848780632
10 0.00452315760776401
11 0.00204328843392432
12 0.00101902172900736
13 0.000446071790065616
14 0.000202305789571255
};
\addlegendentry{OAMP-SA (i)}
\addplot [semithick, color0, mark=*, mark size=3, mark options={solid,fill opacity=0}, only marks]
table {%
5 0.09899
6 0.06104
7 0.03607
8 0.01890
9 0.009388
10 0.0045192
11 0.0021194
12 0.00084453
13 0.00038122
14 0.000092176
};
\addlegendentry{OAMP-SA (ii)}
\addplot [semithick, color3, mark=asterisk, mark size=3, mark options={solid}, only marks]
table {%
5 0.0952115952968597
6 0.0602738857269287
7 0.0378514118492603
8 0.0230371356010437
9 0.0126266283914447
10 0.00769561203196645
11 0.00427739368751645
12 0.00238096434623003
13 0.00131273968145251
14 0.00052724388660863
};
\addlegendentry{OAMPNet2 (i)}
\addplot [semithick, color2, mark=*, mark size=3, mark options={solid,fill opacity=0}, only marks]
table {%
5 0.09959
6 0.0648
7 0.04101
8 0.02388
9 0.01367
10 0.00769561203196645
11 0.0042199
12 0.002541
13 0.0012183
14 0.00044967
};
\addlegendentry{OAMPNet2 (ii)}
\addplot [semithick, color4, mark=asterisk, mark size=3, mark options={solid}, only marks]
table {%
5 0.142420768737793
6 0.112610392272472
7 0.0876428633928299
8 0.0643774420022964
9 0.0455110594630241
10 0.0305594131350517
11 0.019557474181056
12 0.01177794393152
13 0.0067697511985898
14 0.00328159960918128
};
\addlegendentry{MMSE}
\addplot [semithick,sienna1408675, mark=asterisk, mark size=3, mark options={solid}, only marks]
table {%
5 0.122700460255146
6 0.0944103598594666
7 0.0716721937060356
8 0.0516703948378563
9 0.0360446460545063
10 0.0243805944919586
11 0.0158737618476152
12 0.00992097612470388
13 0.00582141149789095
14 0.00300411623902619
};
\addlegendentry{OAMP}
\end{axis}

\end{tikzpicture}
	}
	\caption{Evaluation of different decoding algorithms over SNR for a $16\times8$ QAM 4 MIMO system using the SER metric. Stars mark evaluations for setting (i) (non-generalizing) with $c_r=c_t=0.1$; circles mark evaluations for setting (ii) (generalizing) with varying  data SNR.}
	\label{fig:ser_over_snr}
\end{figure}

In Figure~\ref{fig:ser_over_snr}, we see the Symbol Error Rate (SER) of the model at $c_t=c_r=0.1$, varying the SNR of the MIMO data. Here, the self-attention model manages to substantially outperform OAMPNet2 over the entire SNR range. Interestingly, we also see that for the high SNR region, the OAMP-SA model in setting (ii) outperforms variant (i). This underlines the excellent generalization capabilities of the model:
although our proposed model is more expressive than the alternative hybrid models, it generalizes well when evaluated on scenarios with SNR values not covered in the training data. This also indicates that for data-driven models training data selection is crucial, and sufficiently difficult scenarios need to be covered during training. 
\subsection{Realistic channel data}\label{sec:realistic_channel}
For this evaluation, we use a more realistic scenario, utilizing 3GPP channel realizations generated via a clustered delay line (CDL)-B model~\cite{etsi2018138}. The investigated MIMO system uses $N_r=64$ antennas, with beam selection. After beam selection, the resulting channel matrices contain $\bar{N}_r=16$ beams, for a channel recorded over 200 time steps using 96 subcarriers. The selection procedure randomly chooses $N_t=12$ users out of a total of 100, creating 20 sets of users for training and another 20 sets for the test set. From the training set, we split 20~\% as a validation set, resulting in $16\times 200\times 96$ channel matrices for training, $4\times 200\times 96$ for validation, and $20\times 200\times 96$ matrices for testing. Here, the training and test set use two disjoint sets of users in order to increase the diversity between the training and test set in order to detect overfitting on specific scenarios. Each channel is then evaluated for two symbol transmissions, for training as well as for testing. The transmissions for the training procedure were randomly drawn from an SNR range of 5-20~dB.  
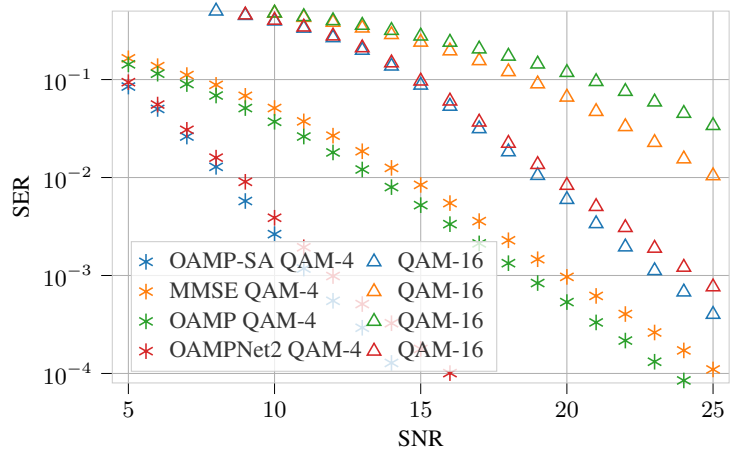
\begin{figure}[h]
	\centering
	\footnotesize
	\resizebox{0.6\linewidth}{!}{
\begin{tikzpicture}

\definecolor{crimson2143940}{RGB}{214,39,40}
\definecolor{darkgray176}{RGB}{176,176,176}
\definecolor{darkorange25512714}{RGB}{255,127,14}
\definecolor{forestgreen4416044}{RGB}{44,160,44}
\definecolor{lightgray204}{RGB}{204,204,204}
\definecolor{steelblue31119180}{RGB}{31,119,180}
\definecolor{mediumpurple148103189}{RGB}{148,103,189}

\begin{axis}[
legend columns=2,
legend cell align={left},
legend style={
  fill opacity=0.8,
  draw opacity=1,
  text opacity=1,
  at={(0.03,0.03)},
  anchor=south west,
  draw=lightgray204
},
height=0.4\columnwidth,
width=0.6\columnwidth,
log basis y={10},
tick align=outside,
tick pos=left,
x grid style={darkgray176},
axis line style={white!80!black},
xlabel={SNR},
xmajorgrids,
xmin=4.45, xmax=25.55,
xtick style={color=black},
y grid style={darkgray176},
ylabel={SER},
ymajorgrids,
ymin=8e-05, ymax=0.5,
ymode=log,
ytick style={color=black},
ytick={1e-06,1e-05,0.0001,0.001,0.01,0.1,1,10},
yticklabels={
  \(\displaystyle {10^{-6}}\),
  \(\displaystyle {10^{-5}}\),
  \(\displaystyle {10^{-4}}\),
  \(\displaystyle {10^{-3}}\),
  \(\displaystyle {10^{-2}}\),
  \(\displaystyle {10^{-1}}\),
  \(\displaystyle {10^{0}}\),
  \(\displaystyle {10^{1}}\)
}
]
\addplot [semithick, steelblue31119180, mark=asterisk, mark size=3, mark options={solid}, only marks]
table {%
5 0.084850437939167
6 0.0501224100589752
7 0.026205500587821
8 0.0128557868301868
9 0.00575857423245907
10 0.0026432282757014
11 0.00117730174679309
12 0.000551648088730872
13 0.000292751850793138
14 0.000128038198454306
15 5.93533004575875e-05
16 2.99479142995551e-05
17 1.33463554448099e-05
18 4.12326426157961e-06
19 2.49565982812783e-06
20 1.41059035740909e-06
21 3.25520829846937e-07
22 0
23 0
24 0
};
\addlegendentry{OAMP-SA QAM-4}
\addplot [semithick, steelblue31119180, mark=triangle, mark size=3, mark options={solid}, only marks]
table {%
5 0.6227787733078
6 0.585530579090118
7 0.54472678899765
8 0.499895513057709
9 0.450582683086395
10 0.395704030990601
11 0.334296226501465
12 0.267266362905502
13 0.198943331837654
14 0.137238994240761
15 0.0876257717609406
16 0.053353413939476
17 0.0314732082188129
18 0.0182929653674364
19 0.0104955499991775
20 0.00596571294590831
21 0.00337771279737353
22 0.00195572921074927
23 0.0011195745319128
24 0.000678276759572327
25 0.000400824676034972
};
\addlegendentry{QAM-16}
\addplot [semithick, darkorange25512714, mark=asterisk, mark size=3, mark options={solid}, only marks]
table {%
5 0.164742603898048
6 0.137035578489304
7 0.11116748303175
8 0.0882984399795532
9 0.0682202577590942
10 0.051232174038887
11 0.0374515391886234
12 0.0267550218850374
13 0.018547972664237
14 0.0125443805009127
15 0.00841243471950293
16 0.00548122776672244
17 0.0035901686642319
18 0.00228906259872019
19 0.00146527786273509
20 0.00096809834940359
21 0.000618706690147519
22 0.0004041884385515
23 0.000261935783782974
24 0.000171875013620593
25 0.0001099174914998
};
\addlegendentry{MMSE QAM-4}
\addplot [semithick, darkorange25512714, mark=triangle, mark size=3, mark options={solid}, only marks]
table {%
5 0.658409178256989
6 0.627735733985901
7 0.593895316123962
8 0.557342052459717
9 0.517825722694397
10 0.47544401884079
11 0.430503785610199
12 0.383840888738632
13 0.335606753826141
14 0.287494391202927
15 0.240599647164345
16 0.196167901158333
17 0.155722036957741
18 0.12024001032114
19 0.0902997031807899
20 0.0661101192235947
21 0.0471753589808941
22 0.0330389514565468
23 0.0227866768836975
24 0.0154873058199883
25 0.0104232830926776
};
\addlegendentry{QAM-16}
\addplot [semithick, forestgreen4416044, mark=asterisk, mark size=3, mark options={solid}, only marks]
table {%
5 0.14301523566246
6 0.114824540913105
7 0.0899306535720825
8 0.0687398537993431
9 0.0511835739016533
10 0.0370966605842113
11 0.0262130666524172
12 0.0180325787514448
13 0.0120789837092161
14 0.00798753183335066
15 0.00522460928186774
16 0.00334255606867373
17 0.00212630280293524
18 0.00133550306782126
19 0.000837456376757473
20 0.000535265018697828
21 0.000331814197124913
22 0.000216688349610195
23 0.000131727414554916
24 8.48523995955475e-05
};
\addlegendentry{OAMP QAM-4}
\addplot [semithick, forestgreen4416044, mark=triangle, mark size=3, mark options={solid}, only marks]
table {%
5 0.643565118312836
6 0.615119159221649
7 0.584018766880035
8 0.55081570148468
9 0.515428960323334
10 0.477940797805786
11 0.438794314861298
12 0.398881137371063
13 0.357737213373184
14 0.317277729511261
15 0.27791440486908
16 0.240273430943489
17 0.205239772796631
18 0.1731136739254
19 0.143999636173248
20 0.118055894970894
21 0.0953469052910805
22 0.0756381303071976
23 0.0590040981769562
24 0.0451422519981861
25 0.0339108109474182
};
\addlegendentry{QAM-16}
\addplot [semithick, crimson2143940, mark=asterisk, mark size=3, mark options={solid}, only marks]
table {%
5 0.0940222293138504
6 0.0555860735476017
7 0.0306913200765848
8 0.0159604996442795
9 0.0090426430106163
10 0.0038853082805872
11 0.00194140546955168
12 0.000987196690402925
13 0.000510958780068904
14 0.000325629371218383
15 0.000177951398654841
16 0.000100802950328216
17 5.47960044059437e-05
18 3.53732721123379e-05
19 1.41058999361121e-05
20 1.08506965261768e-05
21 6.94444406690309e-06
22 4.34027742812759e-06
23 1.95312509276846e-06
24 2.0616321307898e-06
};
\addlegendentry{OAMPNet2 QAM-4}
\addplot [semithick, crimson2143940, mark=triangle, mark size=3, mark options={solid}, only marks]
table {%
5 0.63619863986969
6 0.595178246498108
7 0.551933288574219
8 0.505974471569061
9 0.456583172082901
10 0.403040200471878
11 0.343710124492645
12 0.278690218925476
13 0.211009934544563
14 0.147949069738388
15 0.0965040996670723
16 0.0603290908038616
17 0.0367929674685001
18 0.0223422292619944
19 0.0135732451453805
20 0.00831184536218643
21 0.0050544710829854
22 0.00308333360590041
23 0.00189604982733727
24 0.00121267419308424
25 0.000765407807193696
};
\addlegendentry{QAM-16}
\end{axis}

\end{tikzpicture}
	}
	\caption{Algorithm comparison on a 3GPP channel scenario after beam selection. The system uses $\bar{N}_r=16$ beams, for $N_t=12$ users and QAM-4 (stars) as well as QAM-16 (triangles) modulation.}
	\label{fig:realistic_over_snr}
\end{figure}
Figure~\ref{fig:realistic_over_snr} shows, that our proposed OAMP-SA model can also improve performance on realistic channel data, and does not degrade in performance for the SNR region not covered in the training distribution ranging from 20-25~dB SNR. Interestingly, OAMP performance for the QAM-16 system degrades compared to MMSE for the high SNR region, which as a result might also affect OAMPNet2 and OAMP-SA performance.
\subsection{Runtime evaluation}\label{sec:runtime_evaluation}
In this experiment, we evaluate the average runtime of the MMSE, OAMP, OAMPNet2, and OAMP-SA algorithms on a GPU system using an NVIDIA RTX 2080Ti. Here, we investigate two systems of size $N_r=64$, $N_t=32$, and  $N_r=512$, $N_t=256$. These system sizes allow for a better assessment of the algorithm's scalability.
         

\begin{table}[h]
    \centering
    \small
        \caption{Average runtime of the investigated algorithms for a $N_r=64$, $N_t=32$ and a $N_r=512$, $N_t=256$ QAM-4 MIMO system on GPU hardware.}
    \begin{tabular}{c||c|c|c|c}
         System&MMSE&OAMP&OAMPNet2&OAMP-SA\\ \hline
         64$\times$32&1.6 $\mu$s&10.9 $\mu$s&11.9$\mu$s&32.8$\mu$s\\
         512$\times$256&47.0$\mu$s&93.8$\mu$s&125.0$\mu$s&237.5$\mu$s
         
    \end{tabular}

    \label{tab:runtime_table}
\end{table}
In Table~\ref{tab:runtime_table}, we see that the OAMP-SA model takes three times as long as OAMPNet2 per channel evaluation for the smaller $64\times32$ system, and about two times as long for the larger $512\times 256$ system. These results show, that while the self-attention model does add some complexity, it scales well to larger systems.   
\section{Conclusion}\label{sec:conclusion}
In this paper, we proposed a self-attention model to enhance the OAMP algorithm for a realistic MIMO system. The introduced deep learning architecture is able to incorporate an analytical prior that makes it more data efficient and improves generalization at limited computational overhead. In our experiments, we showed that our model is able to generalize well across a wide SNR range, and is able to outperform existing OAMP-based approaches on realistic channel data while scaling well to large MIMO systems. For future investigations, we would like to explore the applicability of self-attention models for other iterative decoding algorithms. 
\bibliographystyle{IEEEbib}
\bibliography{IEEEabrv,mimo_bib}

\end{document}